\documentclass[prl,amsmath,amssymb,onecolumn]{revtex4}
\usepackage{mathrsfs}
\usepackage{amssymb}

\input epsf.tex
\usepackage{graphicx}
\usepackage{amsthm}

\begin{document}

\title{Security of quantum key distribution with state-dependent imperfections}

\author{Hong-Wei Li$^{1,2,*}$,  Zhen-Qiang Yin$^{1,a}$, Shuang Wang$^{1}$, Wan-Su Bao$^{2}$,
Guang-Can Guo$^1$, Zheng-Fu Han$^{1,b}$ }

 \affiliation
 {$^1$ Key Laboratory of Quantum Information,University of Science and Technology of China,Hefei, 230026,
 China\\$^2$ Zhengzhou Information Science and Technology Institute, Zhengzhou, 450004,
 China}

 \date{\today}
\begin{abstract}
In practical quantum key distribution system, the state preparation
and measurement have state-dependent imperfections comparing with
the ideal BB84 protocol. If the state-dependent imperfection can not
be regarded as an unitary transformation, it should not be
considered as part of quantum channel noise introduced by the
eavesdropper, the commonly used secret key rate formula GLLP can not
be applied correspondingly. In this paper, the unconditional
security of quantum key distribution with state-dependent
imperfections will be analyzed by estimating upper bound of the
phase error rate in the quantum channel and the imperfect
measurement. Interestingly, since Eve can not control all phase
error in the quantum key distribution system, the final secret key
rate under constant quantum bit error rate can be improved comparing
with the perfect quantum key distribution protocol.
\end{abstract}
\maketitle

\section{Introduction}\label{Introduction}
Quantum key distribution (QKD) is the art of sharing secret keys
between the transmitter Alice and receiver Bob.
 It has unconditional security, even if unlimited computational power and storage capacity are controlled by the eavesdropper Eve.
Since the QKD protocol has been proposed by Bennett and Brassard in
1984 \cite{review BB84}, the unconditional security attracts a lot
of attentions both in theory and experimental sides  \cite{review
Sc1}. Theoretical physicists have analyzed unconditional security of
QKD in many respects. Initially, Lo and Chau \cite{review Lo}
proposed the security analysis with the help of quantum computer.
Then, Shor and Preskill \cite{review Sh} proved security of
prepare-and-measure protocol is equival to entanglement-based
protocol, thus unconditional security of QKD has been proved
combining with the CSS code and entanglement distillation and
purification (EDP) technology. Without applying the EDP technology,
Renner \cite{review Re} has analyzed security of QKD with
information theory method. More recently, Horodecki et al.
\cite{review Horodecki} have analyzed security of QKD based on
Private-entanglement states. Inspired by Horodecki's mind, Renes and
Smith \cite{review Smith} have analyzed noisy processing allows some
phase errors to be left uncorrected without compromising
unconditional security of the key. However, all of the security
analysis are based on perfect states preparation and measurement.
The first unconditional security of QKD based on imperfect devices
was proposed by Gottesman, Lo, Lukenhaus, and Preskill \cite{review
GLLP} (GLLP formula), they proved that only the single photon state
transmitted in the quantum channel can be used to generate the final
secret key. Applying the GLLP formula and decoy state method
\cite{review Hwang}, security of the decoy state QKD has been
analyzed by Lo \cite{review Lo2} and Wang \cite{review Wang}
respectively. Correspondingly, the secret key transmission distance
can be improved greatly with decoy state method \cite{decoy
experiment1,decoy experiment2}. More recently, Berta et al.
\cite{review RR1, review RR2} have given a method for proving Bob's
device independent QKD protocol by using the uncertainty relation,
which is related to the earlier work by Koashi \cite{Koashi}, but it
also requires that the state preparation in Alice's side should be
well characterized.

Obviously, if the imperfection is basis-dependent, we can consider a
slightly changed protocol, where the state preparation and
measurement are perfect, while there is an virtual unitary
transformation controlled by Eve introduces the basis-dependent
imperfection in the quantum channel. Since security of the original
protocol is no less than the slightly changed protocol, the final
secret key rate can be estimated utilizing the GLLP formula.
However, most of the imperfection in states preparation and
 measurement are state-dependent \cite{lar1,lar2}, which can not be controlled by Eve in the security analysis. For instance, the wave plate may be inaccurate
  in polarization based QKD system, while the phase modulator may be
  modulated by inaccurate voltage in phase-coding QKD system. If the
  imperfection can not be illustrated as an unitary transformation, it can no be considered as part of the quantum channel controlled by Eve.

   In this paper, security of practical QKD system with state-dependent imperfections will be
  analyzed by considering imperfect states preparation in Alice's
  side and imperfect states measurement in Bob's side respectively.
We apply the EDP technology by considering the most general
imperfection, and a much better secret key rate under constant
imperfect parameters has been analyzed in comparation with previous
works. Comparing with the security analysis given by Mar\o y et al.
\cite{lar1} and Lydersen et al. \cite{lar2}, we apply a much simpler
method and get a much higher secret key rate. We consider that
states prepared by Alice and measured by Bob both have individual
imperfections. The whole security analysis can be divided into two
steps based on an virtual protocol. Firstly, we consider that Alice
prepares perfect entangled quantum state pairs, and she keeps half
of the perfect entangled quantum state, sends half of the imperfect
modulated quantum state to Bob, which illustrate the imperfect
states preparation. Meanwhile, Bob
 applies perfect Hadamard transformation in the receiver's side,
 thus
Alice and Bob can share the maximally entangled quantum state
utilizing the EDP technology. Secondly, Alice applies perfect
measurement with her maximally entangled quantum states, and Bob
applies imperfect measurement with his entangled quantum states
correspondingly, finally they can establish the raw key. Similar to
Shor and Preskill's \cite{review Sh} security analysis, security of
the practical QKD is equal to the virtual protocol with the EDP
technology and imperfect measurement. Since the phase error
introduced by Bob's imperfect measurement should not be controlled
by Eve, we can get a much higher secret key rate correspondingly.
The similar result has also been given by Renner et al.
\cite{renner1,renner2,renner3}, they proved that adding noise in the
classical post processing can improve the secret key rate by
considering that phase errors introduced in the post processing can
not be controlled by the eavesdropper Eve \cite{review Smith}.
Comparing with the security analysis given by Renner et al., the
noise introduced by the imperfect device are precisely known by Eve,
since the random encoding choice, the imperfection can not be
corrected or controlled by Eve. Thus, the exactly known but can not
be controlled imperfection is similar to adding noise as the
security analysis model given by Kraus et al.
\cite{renner1,renner2}.

\section {Security of quantum key distribution with perfect states preparation and measurement}

Before introducing the method to analyze security of QKD with
imperfect devices, security of QKD with perfect devices will be
analyzed in this section. Suppose that Alice and Bob choose the
polarization encoding QKD system in our security analysis, the
standard prepare-and-measure QKD protocol will be introduced in the
following section. In Alice's side, the classical bit 0 is randomly
encoded by quantum states $ |0\textordmasculine \rangle$ or $
|45\textordmasculine \rangle$, the classical bit 1 is randomly
encoded by quantum states $ |90\textordmasculine \rangle$ or $
|-45\textordmasculine \rangle$. In Bob's side, he randomly choose
rectilinear basis $\{ | 0\textordmasculine\rangle, |
90\textordmasculine\rangle\}$ or diagonal basis $\{ |
45\textordmasculine \rangle, | -45\textordmasculine\rangle\}$ to
measure the quantum state transmitted through the quantum channel.
After Bob's perfect measurement and some classical steps of QKD
(sifting, parameter estimation, error correction and privacy
amplification), secret key bits can be shared between Alice and Bob.

Following the technique obtained by Shor and Preskill \cite{review
Sh}, security of prepare-and-measure QKD protocol is equal to
security of entanglement-based QKD protocol, which can be
constructed by considering the corresponding prepare-and-measure
encoding scheme as shown in Fig.1.

\begin{figure}[!h]\center
\resizebox{10cm}{!}{
\includegraphics{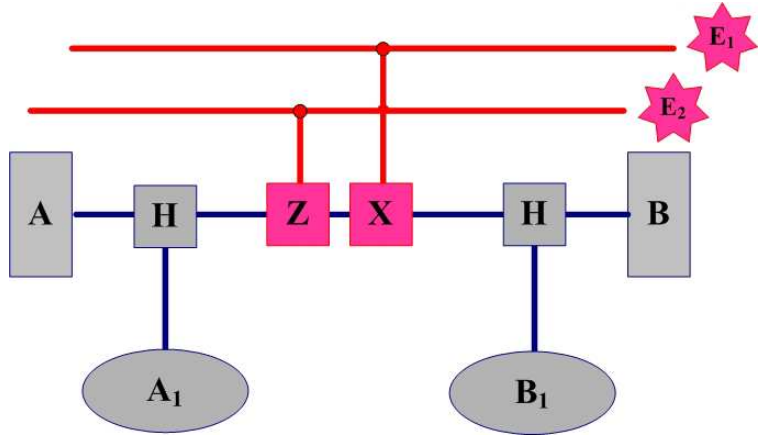}}
\caption{Entanglement-based protocol with Pauli channel and
eavesdropper Eve. $Z$ is Eve's phase error operation, $X$ is Eve's
bit error operation. $A_1$ is part of Alice's system, $B_1$ is part
of Bob's system.}
\end{figure}
 Alice prepares maximally entangled pairs
$|\phi_1\rangle=\frac{1}{\sqrt{2}}(|00\rangle_{AB}+|11\rangle_{AB})$.
After applying the Hadamard operation randomly to the second part of
the pair, she sends Bob half of the pair. Bob acknowledges the
reception of his state and applies the Hadamard operation randomly.
In the security analysis, the most generally noisy channels we need
to consider are Pauli channels. By considering Eve's eavesdropping
in the Pauli channel, the quantum state about Alice, Bob and Eve is
given by

\begin{equation}
\begin{array}{lll}
\sum\limits_{u,v,i,j}\sqrt{P_{uv}Q_{ij}}(I_A\bigotimes
H_{B_1}^{i}X_{E_1}^{u}Z_{E_2}^{v}H_{A_1}^{j}|\phi_1\rangle|u\rangle_{E_1}|v\rangle_{E_2}|i\rangle_{B_1}|j\rangle_{A_1}),
\end{array}
\end{equation}
where $ H= \frac{1}{\sqrt{2}}\left (
\begin{array}{ccc}
 1 & 1  \\
 1 & -1  \\
\end{array}
\right) $ is the perfect Hadamard operator, which means the
transformation between different bases in practical QKD system. $
\left (
\begin{array}{ccc}
 0 & 1  \\
 1 & 0  \\
\end{array}
\right) $ is the X operator, which means the bit error introduced by
Eve. $ \left (
\begin{array}{ccc}
 1 & 0  \\
 0 & -1  \\
\end{array}
\right) $ is the Z operator, which means the phase error introduced
by Eve. Correspondingly, $XZ$ means the bit phase error introduced
by Eve in the quantum channel. $P_{uv}$, $u,v \in\{0,1\}$ means the
probability of the $X^{u}Z^{v}$ operator introduced by Eve, which
should be normalized by the following equation

\begin{equation}
\begin{array}{lll}
\sum\limits_{u,v}P_{uv}=1.
\end{array}
\end{equation}
$Q_{ij}$, $i,j \in\{0,1\}$means the probability of $H^{i}$ and
$H^{j}$ matrix introduced by Alice and Bob respectively, which
satisfies $Q_{ij}=\frac{1}{4}$ for Alice and Bob's randomly choice.

After the sifting step, the case of $i\neq j$ will be discarded. We
trace out $A_1$, $B_1$ and Eve's systems to get the following
equation

\begin{equation}                           \label{practical state}
\begin{array}{lll}
\rho_{AB}=\sum\limits_{u,v}P_{uv}(\frac{1}{2}I_A\bigotimes
X_{E_1}^{u}Z_{E_2}^{v}|\phi_1\rangle\langle\phi_1|Z_{E_2}^{v}X_{E_1}^{u}\bigotimes I_A\\
+\frac{1}{2}I_A\bigotimes
H_{B_1}X_{E_1}^{u}Z_{E_2}^{v}H_{A_1}|\phi_1\rangle\langle\phi_1|H_{A_1}Z_{E_2}^{v}X_{E_1}^{u}H_{B_1})
\bigotimes I_A.
\end{array}
\end{equation}

There are bit errors and phase errors in the Pauli channel, all of
errors are considered to be introduced by Eve in the security
analysis. After transmitting through the quantum channel, the
initially shared maximally entangled state can be transformed into
Bell states as the following equation

\begin{equation}               \label{Bell states}
\begin{array}{lll}
|\phi_1\rangle=\frac{1}{\sqrt{2}}(|00\rangle_{AB}+|11\rangle_{AB}),\\
|\phi_2\rangle=\frac{1}{\sqrt{2}}(|01\rangle_{AB}+|10\rangle_{AB}),\\
|\phi_3\rangle=\frac{1}{\sqrt{2}}(|00\rangle_{AB}-|11\rangle_{AB}),\\
|\phi_4\rangle=\frac{1}{\sqrt{2}}(|01\rangle_{AB}-|10\rangle_{AB}).
\end{array}
\end{equation}

If the maximally entangled pairs $|\pi_1\rangle$ is transformed into
Bell state $|\phi_1\rangle$, there is no error can be introduced in
the quantum channel. However, if the maximally entangled pairs
$|\phi_1\rangle$ is transformed into Bell states $|\phi_2\rangle$,
$|\phi_3\rangle$ and $|\phi_4\rangle$ respectively, the bit error,
phase error and bit phase error will be introduced by Eve
correspondingly. Thus, the bit error rate and phase error rate can
be given by

\begin{equation}                    \label{error rate}
\begin{array}{lll}
e_{bit}&=\langle\phi_2|\rho_{AB}|\phi_2\rangle+\langle\phi_4|\rho_{AB}|\phi_4\rangle,\\
e_{phase}&=\langle\phi_3|\rho_{AB}|\phi_3\rangle+\langle\phi_4|\rho_{AB}|\phi_4\rangle.
\end{array}
\end{equation}

The bit error rate and phase error rate should be calculated when we
analyze unconditional security of QKD. In practical QKD system,
quantum bit error rate can be estimated after the
 parameter estimation step in the classical part of QKD protocol. The
 main difficulty in security
 analysis is how to estimate upper bound of the phase error rate.

Combining equations (\ref{practical state}), (\ref{Bell states})
with (\ref{error rate}), the phase error rate minus the bit error
rate is

\begin{equation}
\begin{array}{lll}
e_{phase}-e_{bit}=\langle\phi_2|\rho_{AB}|\phi_2\rangle-\langle\phi_3|\rho_{AB}|\phi_3\rangle=0.
\end{array}
\end{equation}
Thus, the phase error can be estimated by the bit error rate
accurately in the perfect device case. Correspondingly, the final
secret key rate can be given by

\begin{equation}                    \label{key}
\begin{array}{lll}
R=1-h(e_{phase})-h(e_{bit})=1-2h(e_{bit}).
\end{array}
\end{equation}
where, $h$ is the binary entropy function. The maximal tolerated bit
error rate in the quantum channel is $0.11$ with equation
(\ref{key}), which has also been given by Shor and Preskill.

\section {Security of quantum key distribution with state-dependent imperfections}

Since practical QKD devices always have some flaws, the photon state
preparation and measurement are always imperfect in practical QKD
realizations. In the most general case, the imperfection is
state-dependent. For example, the deflection angle has slight
differences between different wave plates in polarization based QKD
system. Similar to the security analysis of QKD with perfect
devices, we will give the security analysis of QKD with imperfect
devices in this section by utilizing the EDP technology and
imperfect measurement. We firstly give the model description about
the imperfect states preparation and measurement, then we will prove
that the imperfect measurement is equal to the perfect measurement
adding the noisy processing in our security analysis, finally
security of the virtual protocol will be analyzed combining with the
imperfect measurement and EDP technology.

\subsection{Device-independent imperfections description}

Angular deviation of the practical device can be used for
illustrating the state-dependent imperfection. In Alice's side, the
classical bit 0 is randomly encoded by quantum states $
|\alpha_1\textordmasculine \rangle$ or $
|45+\alpha_2\textordmasculine \rangle$, while the classical bit 1 is
randomly encoded by quantum states $ |90+\alpha_3\textordmasculine
\rangle$ or $ |-45+\alpha_4\textordmasculine \rangle$, where
$\alpha_1$, $\alpha_2$, $\alpha_3$ and $\alpha_4$ are security
parameters for illustrating Alice's angular deviations. In Bob's
side, he randomly choose the imperfect rectilinear basis $\{ |
\beta_1\textordmasculine\rangle, |
90+\beta_3\textordmasculine\rangle\}$ or the imperfect diagonal
basis $\{ | 45+\beta_2\textordmasculine\rangle, |
-45+\beta_4\textordmasculine\rangle\}$ to measure the quantum state
transmitted in the quantum channel, where $\beta_1$, $\beta_2$,
$\beta_3$ and $\beta_4$ are security parameters for illustrating
Bob's angular deviation. Since the random encoding and decoding
choice, all of the imperfection can not be controlled or corrected
by the eavesdropper, detailed illustration of the imperfect
parameter can be given as in Fig. 2.

\begin{figure}[!h]\center
\resizebox{10cm}{!}{
\includegraphics{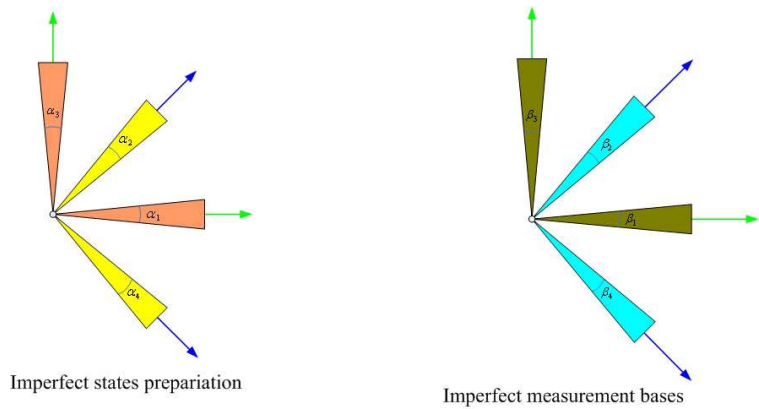}}
\caption{The most general imperfect states preparation and
measurement in practical QKD experimental realization, where
$\alpha_1$, $\alpha_2$, $\alpha_3$ and $\alpha_4$ illustrates the
imperfect states preparation, $\beta_1$, $\beta_2$, $\beta_3$ and
$\beta_4$ illustrates the imperfect measurement.}
\end{figure}
If the security parameter can be satisfied with $\alpha_1=\alpha_3$,
$\alpha_2=\alpha_4$, $\beta_1=\beta_3$
 and $\beta_2=\beta_4$, it will be the basis-dependent imperfection
 correspondingly.

\subsection{Imperfect measurement}

In practical QKD system with imperfect devices as illustrated in the
previous subsection, Bob gets the classical bit 0 with the
projective measurement operator
$|\beta_1\textordmasculine\rangle\langle\beta_1\textordmasculine|$
and
$|45+\beta_2\textordmasculine\rangle\langle45+\beta_2\textordmasculine|$,
gets the classical bit 1 with the projective measurement
$|90+\beta_3\textordmasculine\rangle\langle90+\beta_3\textordmasculine|$
and
$|-45+\beta_4\textordmasculine\rangle\langle-45+\beta_4\textordmasculine|$
respectively. Since the rectilinear basis and diagonal basis will be
selected randomly, the quantum bit error rate introduced by the
imperfect measurement can be given by
\begin{equation}                    \label{bit error rate2}
\begin{array}{ll}
e_{bit1}\\
=
\frac{1}{2}[\frac{1}{2}(\frac{\langle90+\beta_3\textordmasculine|0\textordmasculine\rangle\langle0\textordmasculine|90+\beta_3\textordmasculine\rangle}{\langle90+\beta_3\textordmasculine|0\textordmasculine\rangle\langle0\textordmasculine|90+\beta_3\textordmasculine\rangle+\langle\beta_1\textordmasculine|0\textordmasculine\rangle\langle0\textordmasculine|\beta_1\textordmasculine\rangle}
+\frac{\langle\beta_1\textordmasculine|90\textordmasculine\rangle\langle90\textordmasculine|\beta_1\textordmasculine\rangle}{\langle\beta_1\textordmasculine|90\textordmasculine\rangle\langle90\textordmasculine|\beta_1\textordmasculine\rangle+\langle90+\beta_3\textordmasculine|90\textordmasculine\rangle\langle90\textordmasculine|90+\beta_3\textordmasculine\rangle})+
\\
 ~~~~~~
\frac{1}{2}(\frac{\langle-45+\beta_4\textordmasculine|45\textordmasculine\rangle\langle45\textordmasculine|-45+\beta_4\textordmasculine\rangle}{\langle-45+\beta_4\textordmasculine|45\textordmasculine\rangle\langle45\textordmasculine|-45+\beta_4\textordmasculine\rangle+\langle45+\beta_2\textordmasculine|45\textordmasculine\rangle\langle45\textordmasculine|45+\beta_2\textordmasculine\rangle}
\\~~~~~+\frac{\langle45+\beta_2\textordmasculine|-45\textordmasculine\rangle\langle-45\textordmasculine|45+\beta_2\textordmasculine\rangle}{\langle45+\beta_2\textordmasculine|-45\textordmasculine\rangle\langle-45\textordmasculine|45+\beta_2\textordmasculine\rangle+\langle-45+\beta_4\textordmasculine|-45\textordmasculine\rangle\langle-45\textordmasculine|-45+\beta_4\textordmasculine\rangle})]
\\
=\frac{1}{2}[\frac{1}{2}(\frac{sin^2{\beta_1}}{sin^2{\beta_1}+cos^2{\beta_3}}+\frac{sin^2{\beta_3}}{sin^2{\beta_3}+cos^2{\beta_1}})+
\frac{1}{2}(\frac{sin^2{\beta_2}}{sin^2{\beta_2}+cos^2{\beta_4}}+\frac{sin^2{\beta_4}}{sin^2{\beta_4}+cos^2{\beta_2}})].
\end{array}
\end{equation}
From this calculation, we can get the result that the imperfect
measurement will introduce bit flipping with the probability
$e_{bit1}$. Comparing with the imperfect measurement, the perfect
measurement will introduce the bit flipping with zero probability.
In our security analysis, Alice and Bob should establish the
maximally entangled pairs before applying the measurement, which
means that the eavesdropper can only get the error bit information
about the secret key through the imperfect measurement in Bob's
side. Thus, we can simplify the imperfect measurement as the perfect
measurement adding a noisy processing protocol, where the bit 0(1)
will be transformed into 1(0) with the probability $e_{bit1}$.

\subsection{Virtual EDP protocol}

We propose the virtual protocol based on the EDP technology as in
Fig. 3.

\begin{figure}[!h]\center
\resizebox{10cm}{!}{
\includegraphics{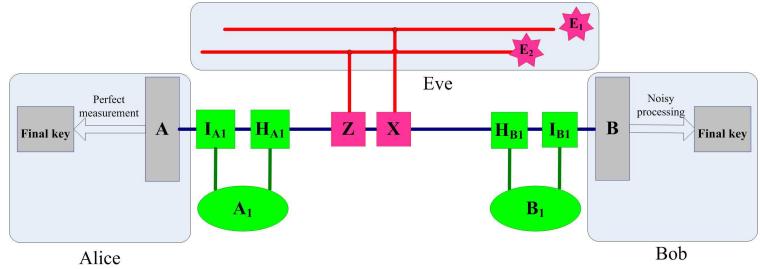}}
\caption{Entanglement-based quantum key distribution protocol with
imperfect devices. We introduce the third party $A_1$, $B_1$ in the
new protocol, which can not be controlled by Alice, Bob and Eve
respectively. In the first step, Alice and Bob share the maximally
entangled pairs. In the second step, Alice applies the perfect
measurement, Bob applies the perfect measurement and a noisy
processing protocol to get the raw key.}
\end{figure}

The new protocol mainly contain two steps: the first step is
considering the maximally entangled state $|\phi_1\rangle$ to be
shared between Alice and Bob. In the rectilinear basis case, the
classical bit 0 is prepared by the quantum state
$cos(\alpha_1)|0\textordmasculine\rangle+sin(\alpha_1)|1\textordmasculine\rangle$,
while the classical bit 1 is prepared by the quantum state
$-sin(\alpha_3)|0\textordmasculine\rangle+cos(\alpha_3)|1\textordmasculine\rangle$.
In the diagonal basis, the classical bit 0 is prepared by the
quantum state
$cos(\alpha_2+\frac{\pi}{4})|0\textordmasculine\rangle+
sin(\alpha_2+\frac{\pi}{4})|1\textordmasculine\rangle$, while the
classical bit 1 is prepared by the quantum state
$cos(\alpha_4-\frac{\pi}{4})|0\textordmasculine\rangle+sin(\alpha_4-\frac{\pi}{4})|1\textordmasculine\rangle$.
For simplicity, the state preparation can also be illustrated as the
following case, Alice prepares the quantum state
$\frac{1}{\sqrt{2}}(|0\textordmasculine\rangle
I_{A_1}|0\textordmasculine\rangle +|1\textordmasculine\rangle
I_{A_1}|1\textordmasculine\rangle)|e_0\rangle+\frac{1}{\sqrt{2}}(|0\textordmasculine\rangle
H_{A_1}|0\textordmasculine\rangle +|1\textordmasculine\rangle
H_{A_1}|1\textordmasculine\rangle)|e_1\rangle$ and transmits half of
the perfect state to Bob, where $H_{A_1}=\left (
\begin{array}{ccc}
 cos(\alpha_2+\frac{\pi}{4}) & cos(\alpha_4-\frac{\pi}{4})  \\
 sin(\alpha_2+\frac{\pi}{4}) & sin(\alpha_4-\frac{\pi}{4})  \\
\end{array}
\right)$, $I_{A_1}= \left (
\begin{array}{ccc}
 cos(\alpha_1) & -sin(\alpha_3)  \\
 sin(\alpha_1) & cos(\alpha_3)  \\
\end{array}
\right)$, $|e_0\rangle$ and $|e_1\rangle$ are Alice's auxiliary
quantum states. If Alice want to transmit the state
$|0\textordmasculine \rangle$ to Bob, the auxiliary quantum state
$|e_0\rangle$ will be selected, and the practical quantum state
$I_{A_1}|0\textordmasculine
\rangle=cos(\alpha_1)|0\textordmasculine\rangle+sin(\alpha_1)|1\textordmasculine\rangle$
will be transmitted in the quantum channel. Since all of the
imperfect state can be analyzed similarly, the non-unitary matrix
$H_{A_1}$ and $I_{A_1}$ can be used for illustrating the imperfect
state preparation. In the receiver's side, Bob applies the unitary
transformation $H_{B_1}$ or $I_{B_1}$ randomly, where $ H_{B_1}=
\frac{1}{\sqrt{2}}\left (
\begin{array}{ccc}
 1 & 1  \\
 1 & -1  \\
\end{array}
\right) $ is the perfect Hadamard transformation, $I_{B_1}= \left (
\begin{array}{ccc}
 1 & 0  \\
 0 & 1  \\
\end{array}
\right)$ is the perfect identity transformation. Since Alice and Bob
can apply the EDP technology, they will share the maximally
entangled quantum pairs before the imperfect measurement.

The second step is applying the perfect and imperfect measurement in
Alice's side and Bob's side respectively. More precisely, Alice
measure the entangled quantum state with perfect rectilinear basis $
\{0\textordmasculine \rangle, |90\textordmasculine \rangle\}$ or
diagonal basis $ \{45\textordmasculine \rangle,
|-45\textordmasculine \rangle\}$. Bob measure the entangled quantum
state with the imperfect rectilinear basis $
\{|\beta_1\textordmasculine \rangle, |90+\beta_3\textordmasculine
\rangle\}$ or imperfect diagonal basis $\{ |
45+\beta_2\textordmasculine\rangle, |
-45+\beta_4\textordmasculine\rangle\}$ correspondingly, then Alice
and Bob will share the raw key.

Similar to the security analysis based on the prefect device, $A_1$
and $B_1$ can not be changed by Alice and Bob in our security
analysis, it can not be changed by Eve simultaneously. However,
$A_1$ and $B_1$ are permitted to share the imperfection information
with Alice, Bob and Eve. In the virtual protocol, the state
preparation and measurement is the same as the original practical
QKD system. If the unconditional security of the virtual protocol
can be proved, security of the practical QKD system can be proved
naturally. By considering Eve's eavesdropping in the Pauli channel,
the quantum state about Alice and Bob before the measurement can be
given by

\begin{equation}
\begin{array}{lll}
\sum\limits_{u,v,i,j}\sqrt{P_{uv}Q_{ij}}(I_A\bigotimes
I_{B_1}^{i+1}H^{i}_{B_1}X_{E_1}^{u}Z_{E_2}^{v}H^{j}_{A_1}I_{A_1}^{j+1}|\phi_1\rangle|u\rangle_{E_1}|v\rangle_{E_2}|i\rangle_{B_1}|j\rangle_{A_1}).
\end{array}
\end{equation}
After the sifting step, the case of $i\neq j$ will be discarded. We
trace out Eve,  $A_1$ and $B_1's$ systems, the density matrix about
Alice and Bob can be given by

\begin{equation}                           \label{imperfect state2}
\begin{array}{lll}
\rho^{'}_{AB}=\sum\limits_{u,v}P_{uv}(\frac{1}{2}I_A\bigotimes
I_{B_1}X_{E_1}^{u}Z_{E_2}^{v}I_{A_1}|\phi_1\rangle\langle\phi_1|I_{A_1}Z_{E_2}^{v}X_{E_1}^{u}I_{B_1}\bigotimes
I_A +
\\\frac{1}{2}I_A\bigotimes
H_{B_1}X_{E_1}^{u}Z_{E_2}^{v}H_{A_1}|\phi_1\rangle\langle\phi_1|H_{A_1}Z_{E_2}^{v}X_{E_1}^{u}H_{B_1}\bigotimes
I_A).
\end{array}
\end{equation}
Suppose that Alice prepare maximally entangled quantum states
$|\phi_1\rangle^{\bigotimes N}$ in her side. After the EDP protocol,
Alice and Bob will share maximally entangled quantum states
$|\phi_1\rangle^{\bigotimes M}$, which can be illustrated as the
following equation
\begin{equation}                    \label{bit error rate}
\begin{array}{lll}
M=N(1-h(e_{bit})-h(e_{phase})),\\
e_{bit}=\langle\phi_2|\rho^{'}_{AB}|\phi_2\rangle+\langle\phi_4|\rho^{'}_{AB}|\phi_4\rangle,\\
e_{phase}=\langle\phi_3|\rho^{'}_{AB}|\phi_3\rangle+\langle\phi_4|\rho^{'}_{AB}|\phi_4\rangle,
\end{array}
\end{equation}
where $e_{bit}$ and $e_{phase}$ are the quantum bit error rate and
phase error rate between Alice and Bob by considering the EDP
technology. Since the calculation of $e_{bit}$ and $e_{phase}$ are
much difficulty, we will get the calculation result based on some
special imperfect parameters and the Mathematic software.
Additionally, Bob will apply the imperfect measurement with the
perfect entanglement quantum state as illustrated in the imperfect
measurement subsection, Alice will apply the perfect measurement
with the perfect entanglement quantum state correspondingly.
Considering the virtual protocol, the practical quantum bit error
rate between Alice and Bob can be estimated by

\begin{equation}                    \label{bit error rate3}
\begin{array}{lll}
Q=1-(1-e_{bit1})(1-e_{bit})-e_{bit}e_{bit1},
\end{array}
\end{equation}
this equation means that the practical quantum bit error rate can be
divided into two cases (considering the EDP protocol and the perfect
measurement in Alice's side and imperfect measurement in Bob's side
respectively). The first case is considering the EDP protocol has
the right bit, the measurement protocol has the error bit. The
second case is considering the EDP protocol has the error bit, the
measurement has the right bit respectively. We can estimate Eve's
information through the whole bit error rate and the phase error
rate in the first step. Finally, the secret key rate can be given by

\begin{equation}                    \label{key rate3}
\begin{array}{lll}
R\geq lim_{N\rightarrow\infty}\frac{M(1-h(e_{bit1}))}{N}\\
= (1-h(e_{phase})-h(e_{bit}))(1-h(e_{bit1})).
\end{array}
\end{equation}
the calculation of which is much complicated for the formula has too
many security parameters, we will give some examples to illustrate
how to use this secret key rate formula in practical QKD system.

 We
give a simple example in the following, we suppose that the
imperfect parameters in our security analysis are
$\alpha_1=\alpha_3=\beta_1=\beta_3=0,\alpha_2=\beta_2=\frac{-\pi}{4},
\alpha_4=\beta_4=\frac{3\pi}{4} $. Thus, we can get
$H_{A_1}=I_{A_1}= \left (\begin{array}{ccc}
 1 & 0  \\
 0 & 1  \\
\end{array}
\right)$, this case means that Alice only send the rectilinear basis
$\{ | 0\textordmasculine\rangle, | 90\textordmasculine\rangle\}$,
and Bob will only measure in the rectilinear basis correspondingly.
The quantum bit error rate and phase error rate in the EDP protocol
can be calculated respectively as the following equation
\begin{equation}
\begin{array}{lll}
e_{bit}=\frac{1}{4}(p_{00}+p_{01}+3p_{10}+3p_{11}),\\
 e_{phase}=\frac{1}{4}(p_{00}+3p_{01}+p_{10}+3p_{11}).
\end{array}
\end{equation}
Correspondingly, upper bound of the phase error rate can be
estimated by
\begin{equation}
\begin{array}{lll}
e_{phase}\leq e_{bit}+\frac{1}{2}\mid p_{10}-p_{01}\mid\leq
e_{bit}+\frac{1}{2}.
\end{array}
\end{equation}
Finally, we can only get zero secret key rate utilizing equation
(\ref{key rate3}). In practical experimental realization, Eve can
measure Alice's states in the same rectilinear basis with 0 bit
error, and she will introduce the perfect man-in-the-middle attack
without being discovered.

\section {Calculation}\label{simulation}

To compare our security analysis with GLLP's security analysis, we
will give the calculation result by considering the case which can
not be analyzed by the GLLP formula in this section. We consider
that the device has individual imperfections both in the
transmitter's side and receiver's side respectively, which means
 quantum states in the same basis maybe have the different
angular deviation. Precisely, we assume that the security parameters
can be satisfied with $\alpha_1=\beta_1=\beta_2=a,
\alpha_2=\alpha_3=\alpha_4=\beta_3=\beta_4=0$. After some lengthy
but not very interesting algebra, the bit error rate and phase error
rate in the first step (Alice and Bob establish the maximally
entangled quantum pairs with the EDP technology) can be calculated
respectively as following equations,
\begin{equation}
\begin{array}{lll}            \label{bit error}
e_{bit} =
\frac{1}{8}[cos^2(a)(p_{11}+p_{10})+sin^2(a)(p_{00}+p_{01})
+cos(a)(2p_{10}-2p_{11})+4p_{01}+p_{10}+p_{11}]\\
~~~~~~~+ \frac{1}{8}[cos^2(a)(p_{11}+p_{10})+sin^2(a)(p_{00}+p_{01})
+cos(a)(-2p_{10}+2p_{11})+p_{10}+5p_{11}],
\end{array}
\end{equation}
\begin{equation}
\begin{array}{lll}             \label{phase error}
e_{phase}=\frac{1}{8}[cos^2(a)(p_{00}+p_{01})+sin^2(a)(p_{10}+p_{11})
+cos(a)(2p_{01}-2p_{00})+4p_{10}+p_{01}+p_{00}]\\
~~~~~~~~~+
\frac{1}{8}[cos^2(a)(p_{11}+p_{10})+sin^2(a)(p_{00}+p_{01})
+cos(a)(-2p_{10}+2p_{11})+p_{10}+5p_{11}].
\end{array}
\end{equation}
Equations (\ref{bit error}) and (\ref{phase error}) can be directly
calculated combining equation (\ref{bit error rate}) with practical
imperfect parameters. From this calculation, we can find that the
phase error rate is equal to the bit error rate in case of all
imperfect parameters are equal to zero.

From this calculation result, upper bound of the phase error rate
can be estimated by considering the following inequation
\begin{equation}
\begin{array}{lll}                         \label{}
|e_{phase}-e_{bit}|
  \\\leq
\frac{1}{8}|[cos^2(a)(p_{11}+p_{10}-p_{00}-p_{01})
 +sin^2(a)(p_{01}+p_{00}-p_{11}-p_{10})\\~~+
 2cos(a)(p_{10}-p_{11}+p_{00}-p_{01})
+3p_{01}-3p_{10}+p_{11}-p_{00}]|\\
=\frac{1}{4}|[cos^2(a)(p_{11}+p_{10}-p_{00}-p_{01})+cos(a)(p_{10}-p_{11}+p_{00}-p_{01})+2p_{01}-2p_{10}]|
\\
 =\frac{1}{4}|[(cos^2(a)-1)(p_{11}+p_{10}-p_{00}-p_{01})+(cos(a)-1)(p_{10}-p_{11}+p_{00}-p_{01})]|
 \\
 \leq\frac{1}{4}(1+sin^2(a)-cos(a))
\end{array}
\end{equation}

\begin{equation}
\begin{array}{lll}                         \label{}
e_{phase}\leq\frac{1}{2}(1+sin^2(a)-cos(a))+e_{bit}.
\end{array}
\end{equation}
Utilizing equation (\ref{bit error rate2}), we can get the bit error
rate $e_{bit1}$ as the following equation

\begin{equation}                    \label{bit error rate2-1}
\begin{array}{lll}
e_{bit1}=\frac{1}{2}\frac{sin^2(a)}{sin^2(a)+1}.
\end{array}
\end{equation}
Combining equations (\ref{bit error rate3}),  (\ref{key rate3}) with
(\ref{bit error rate2-1}),  we can get the final secret key rate
formula as the following equation

\begin{equation}                    \label{key rate3-1}
\begin{array}{lll}
R\geq
(1-h(\frac{Q-e_{bit1}}{1-2e_{bit1}}+\frac{1}{2}(1+sin^2(a)-cos(a)))-h(\frac{Q-e_{bit1}}{1-2e_{bit1}}))(1-h(e_{bit1})),
\end{array}
\end{equation}
combining wit this formula, we give the simulation result of the
final secret key rate by considering practical imperfect security
parameters as in Fig. 4.
\begin{figure}[!h]\center
\resizebox{10cm}{!}{
\includegraphics{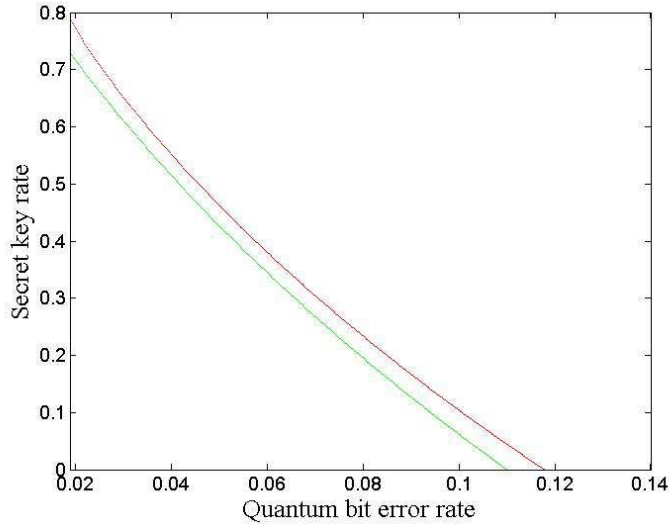}}
\caption{Final secret key rate with perfect and imperfect devices.
The blue line means the perfect devices case, which can be satisfied
with equation (\ref{key}). The red line means security of imperfect
devices by considering parameters $\alpha_1=\beta_1=\beta_2=0.2,
\alpha_2=\alpha_3=\alpha_4=\beta_3=\beta_4=0$.}
\end{figure}

Since the detection setup has the individual imperfection in our
security analysis, the GLLP security analysis result can not be
applied in this case. Comparing with the perfect QKD protocol, the
final secret key rate has been improved in our calculation result,
the reason for which is that Eve can not control the phase error
introduced by Bob's imperfect measurement, and it should no be
corrected by Alice and Bob correspondingly.

\section {conclusions}\label{conclusions}
In practical quantum key distribution realizations, the
state-dependent imperfection in Alice and Bob's side can not be
satisfied with the GLLP formula. A simple security proof of QKD with
state-dependent imperfect states preparation and measurement have
been analyzed in this paper. Our security analysis result shows that
the imperfect QKD system maybe tolerate much higher quantum bit
error rate comparing with the previous security analysis.

\section {Acknowledgements} \noindent The author Hong-Wei Li
would like to thank Lars Lydersen for his helpful discussion and
comments. The author would like to thank the anonymous referees,
they had put a real great effort into reviewing this article, and
they had provided lots of useful feedback that helped to improve the
presentation of this article. This work was supported by National
Fundamental Research Program of China (2006CB921900), National
Natural Science Foundation of China (60537020, 60621064) and the
Innovation Funds of Chinese Academy of Sciences.

\end{document}